\documentstyle[11pt,aas2pp4,tighten]{article}

\lefthead{HALL, OSMER, GREEN, PORTER, WARREN}
\righthead{DEEP MULTICOLOR SURVEY I.}
\slugcomment{Accepted for The Astrophysical Journal Supplement, 1996}

\begin{document}

\title{A DEEP MULTICOLOR SURVEY I. IMAGING OBSERVATIONS AND CATALOG OF STELLAR OBJECTS}

\author{Patrick B. Hall\footnote{Visiting Student, Kitt Peak National Observatory, National Optical Astronomy Observatories, operated by AURA Inc., under contract with the National Science Foundation.}}
\affil{Steward Observatory, University of Arizona, Tucson AZ 85721 \\
E-mail: phall@as.arizona.edu}
\author{Patrick S. Osmer}
\affil{Astronomy Department, The Ohio State University, 174 West 18th Avenue, Columbus, OH 43210 \\
E-mail: posmer@payne.mps.ohio-state.edu}
\author{Richard F. Green, Alain C. Porter\footnote{Deceased.}}
\affil{National Optical Astronomy Observatories, P.O. Box 26732, Tucson AZ 85726 \\
E-mail: rgreen@noao.edu}
\author{Stephen J. Warren}
\affil{Imperial College, Astrophysics Group, The Blackett Laboratory, Prince Consort Road, London SW7 2BZ \\
E-mail: s.j.warren@ic.ac.uk}


\begin{abstract}

We have used the KPNO 4-meter Mayall telescope to image 0.83 square degrees
of sky in six fields at high galactic latitude in six filters spanning
3000-10000\AA\ to magnitude limits ranging from 22.1 to 23.8.
We have assembled a catalog of 21,375 stellar objects detected in the fields 
for use primarily in conducting a multicolor search for quasars.  
This paper describes the data reduction techniques used on the CCD data, 
the methods used to construct the stellar object catalog,
and the simulations performed to understand its completeness and contamination.

\end{abstract}

\section{Introduction}

Quasars, thought to be powered by the accretion of matter onto supermassive
($\sim 10^{7-10}\ \rm{M}_{\sun}$) black holes at the centers of galaxies,
are the most luminous individual objects in the universe (-31$<$M$<$-23).
As such, they can be seen to greater distances than any other objects,
providing us with a unique opportunity to study the universe back 
to times when it was only $\sim$10\% of its current age.
Quasars yield information about the universe through their 
absorption-line systems and associated galaxies (e.g. \cite{ss92}), through 
the galaxies in quasar environments (\cite{ye93}, and references therein), 
and through the lensing of quasars by intervening systems (\cite{kt91}), 
to name but a few.
Additionally, the mere existence of quasars at the observed space densities at 
redshifts z$>$4, when the universe was only $\sim$10\% of its current age, 
constrains models of galaxy and structure formation in the universe 
(e.g. \cite{tur91,loe93,ns93,hr93,ka94}).  

One direct way of using quasars to study the universe is to map the 
evolution of the quasar luminosity function (QLF) with lookback time; that is, 
the number of quasars as a function of absolute magnitude and redshift
(e.g., \cite{bqs,hfc93}).  
Previous studies of the QLF have shown that the space density of the 
bulk of the quasar population, 
with -28$<$M$<$-23, was considerably greater in the past, by a factor of 
$\sim$100 at z=2 relative to the current comoving density.
This was likely due to the evolution of the quasar population in both 
luminosity and in density, though the exact form of the evolution is not well
constrained (\cite{hfc93}).
The redshift distribution of these quasars may peak at z$\sim$2-3 and decrease 
at z$>$3, though not as swiftly as at z$<$2, as first proposed by Osmer (1982).

Recently Warren, Hewett, \& Osmer (1994; hereafter WHO), using a 
multicolor-selected sample of 86 quasars with 2.2$<$z$<$4.5, showed that 
there is indeed a decline in the observed space density of quasars with z$>$3.
This conclusion has been confirmed by the recent results of Schmidt, Schneider,
\& Gunn (1995), using a Ly$\alpha$-selected sample of 90 quasars with 
2.75$<$z$<$4.75.
Although some groups have previously found no evidence for such a decline (e.g.
Koo \& Kron 1988 (KK88), \cite{gv92}), these more recent studies argue that 
the decline is real.
However, it should be noted that WHO cannot rule out that the decline they 
see in the {\it observed} quasar space density at z$>$3 is due to dust 
obscuration, as suggested by e.g. Fall \& Pei (1993), and not reflective
of a decrease in the {\it true} space density of quasars at high redshift.
One of the reasons for this contention over the behavior of the QLF at high 
redshift is that each sample of quasars typically contains only a few such 
objects.  In addition, accurate determination of the space density of quasars
at high redshift requires careful modelling of the selection effects present in
each survey in order to make incompleteness corrections, as discussed e.g. in 
Warren et al. (1991a).

There is obviously a need for additional carefully-selected samples of quasars 
with z$>$2.2.  In 1990 we initiated a deep multicolor survey for quasars with 
the intent of obtaining such a sample.  
Our survey was enabled by the arrival of the first $2048 \times 2048$ CCD at
the KPNO 4-m telescope; the detector and telescope offered at that time a
unique combination of large areal coverage on the sky, spectral sensitivity
across the optical band, and deep limiting magnitude.
The survey was conducted in six optical filters spanning 3000-9000\AA\ in six 
different high-Galactic-latitude fields.  The primary objective was to 
construct a well-defined stellar locus from which quasar candidates could be 
easily isolated via their unusual colors.  However, many other interesting 
questions can be addressed with the survey.  We achieved good photometric 
calibration and the star-galaxy classification effects are well understood, 
and thus the survey should be useful for studies of Galactic star 
counts (\cite{rm93}).  The large number of galaxies detected in the survey also
forms a valuable database for the study of field and emission-line galaxies.  
Our wide wavelength coverage makes the survey especially useful for studies of 
field galaxy color evolution and for the identification of rare galaxy types 
such as E+A galaxies (\cite{liu96}).

Quasars with -25$<$M$<$-23 at z=3-5 are found at apparent magnitudes of roughly
20$<$R$<$23, so our goal was to image to R=24 (R=23.5 was achieved).  
Such a faint limiting magnitude also probes the upper end of the Seyfert 
luminosity function at z$<$3.
To our limiting magnitudes, the survey region contains approximately 30,000 
objects, of which about half are galaxies.  
Based on previous knowledge of the apparent luminosity function of 
quasars (\cite{maj93}), it should contain almost 200 quasars.  
Our ultimate goal is to make available 
catalogs of the quasars, galaxies, and stars containing accurate positions, 
magnitudes, and photometric error estimates for all the objects.

The survey for quasars had two principal objectives: 1) to investigate the 
luminosity function of high-redshift quasars ($z>3$) at fainter magnitudes than
had been done before, for which the absolute magnitudes were more comparable 
with previous surveys at lower redshift; and 2) to determine the luminosity 
function of quasars with $z\leq2.2$ at faint enough magnitudes to constrain 
the debate on luminosity vs. density evolution, which has been unresolved for 
lack of data at the faint end of the luminosity function, where the differences
between the two hypotheses are pronounced.

This paper describes the CCD imaging observations done for the survey and the 
data reduction methods used to construct the catalog of stellar objects from 
which quasar candidates were chosen.  
(We define a {\sl stellar object} as an object detected in our images that is 
found by our classification routines to have an appearance consistent with an 
unresolved point source in a certain minimum number of images, as detailed in 
\S 4.3 below.)
Paper II (\cite{dms2}) discusses quasar candidate selection and initial
spectroscopic results, and makes a comparison with expectations from previous
luminosity function studies.  Paper III (\cite{dms3}) will present final
spectroscopic results and a detailed derivation of the luminosity function.

This paper is organized as follows.  
In \S 2 we describe the observations, the survey fields, and the filters used.
We cover in \S 3 the main features of the data reduction steps.  
In \S 4 we summarize how objects were detected and cataloged using FOCAS and 
how the photometric measurements were made.  
We discuss in \S 5 the photometric calibration and analysis of the errors,
and in \S 6 the astrometric calibrations.
Finally, in \S 7 we summarize the properties of the resulting catalog of 
stellar objects.

\section{Observations}

Images were obtained at the prime focus (f/2.7) of the KPNO 4-meter Mayall 
telescope with a Tektronix 2048x2048 CCD with 27-micron pixels, giving a 
plate scale of 0$\farcs$529/pixel.  
A gain of 11~e-/ADU was used, yielding a full well of $\sim$30,000~ADU.
The particular CCD we used was an engineering grade,
thick chip with poor cosmetics.  
It had a useful area of 1860x2048 pixels due to an excess of variable 
bad columns on one side of the chip, and also had a large number 
of traps evident in the lower light levels of the UBV images.

We used a set of six filters covering the wavelength range from 3000 to 
9000~\AA.  The set consisted of standard KPNO Johnson UBV filters 
($\lambda_{eff}=$3595~\AA, 4364~\AA, and 5563~\AA~respectively) 
plus three special red/near-IR filters: R\arcmin, I75, and I86 
($\lambda_{eff}=$6615~\AA, 7425~\AA, and 8586~\AA~respectively).  
The response curves for each filter in the absence of atmospheric extinction
are plotted in Figure 1.  Each curve is the product of the quantum efficiency 
of the CCD, the transmission of the filter, the transmission of the 4-meter 
prime focus doublet corrector, and one reflection off aluminum.  
The R\arcmin\ filter is narrower than the standard KPNO Harris (not Mould) R 
filter, having less of a red tail and also being less transmissive overall, but
it has very nearly the same effective wavelength (6615~\AA) as the standard 
KPNO Harris R filter (6625~\AA).  Thus for most objects it is safe to treat 
magnitudes in this filter as a standard Harris R filter; only the zeropoint of 
the photometric solution will differ substantially from the solution for a 
Harris R filter (although there will be a color term for very red objects).  
Hereafter the R\arcmin\ filter will be referred to simply as R.  The I75 and 
I86 filters each cover about half the wavelength range of the standard I filter.
These three red filters were designed with narrower than usual bandpasses 
in order to improve our ability to detect quasars with redshifts as 
large as 5.5 and to distinguish z$>$4 quasars from very 
late-type stars via the strong excess from L$\alpha$ emission.  
For z$\leq$5.5, the I86 filter is still longward of
redshifted L$\alpha$ emission, which is normally the strongest line
in quasar spectra, and therefore provides a discriminant between
high-z quasars and later-type stars. 

\begin{figure}[t]
\plotone{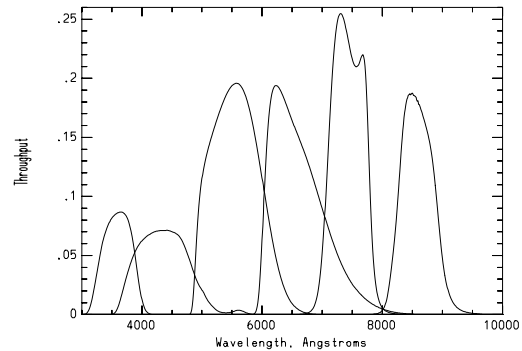}
\caption{
The response curves of the survey in each filter (UBVRI75I86 from left to 
right).  Each curve is the product of the quantum efficiency of the CCD,
the transmission of the filter, the transmission of the 4-meter prime focus 
doublet corrector, and one reflection off aluminum.  Atmospheric extinction
has not been figured into these curves.
}
\end{figure}

Successful 
imaging runs, affected only occasionally by light cirrus, were carried out on 
1990 August 21-24 and 1991 April 14-17, resulting in images being obtained 
in twelve CCD fields.  Our fields are arranged on the sky in six pairs of two 
adjacent individual CCD fields (with slight overlaps),
and were chosen to be at high galactic
latitude away from known dust clouds.  A list of the coordinates,
area, and galactic extinction parameters for each field is given in Table 1.

We attempted to obtain two sky-noise-limited images of each field in every 
filter, offset by typically ten arcseconds from each other, to help reduce 
spurious detections of cosmic ray hits and bad pixels.  
We succeeded in obtaining at least two offset images except
in the 01e field and the 10e field in the B filter.
Where more than two offset images were obtained, the two best in terms of 
seeing, depth, and focus quality across the chip were selected to be the final
two images used in the reduction process. 
Various parameters of these 125 CCD images that comprise the survey
are listed in Table 2.

\section{Data Reduction}

The CCD images in all filters of each field were reduced 
using IRAF\footnote{The Image Reduction and Analysis Facility (IRAF) is
distributed by the National Optical Astronomy Observatories, which is operated
by AURA, Inc., under contract to the National Science Foundation.}.  
The bias level was subtracted from each column 
separately on all field, dark, and domeflat images.  
The appropriate dark current image was subtracted from each field image, and 
domeflats were used to remove the pixel-to-pixel sensitivity variations.  
A separate bad pixel mask needed to be made for each image. 
Each bad area was interpolated across its narrowest dimension.

It was found at this point that the images still showed random, nonrepeatable
gradients and variations in the sky level across the chip.  This was  
particularly true in the three reddest bands, where the variations were
typically 5\% peak-to-peak.  These variations were assumed to be random
additive variations in the sky values, and were removed using a 
two-step process.
First, all images in a particular filter for each run were combined to form 
a ``skyflat'' (identical to the ``supersky'' described in Gullixson 1992)
using a program supplied by Todd Boroson.  
The images were scaled by the mode of a central region and then median filtered to 
form the skyflat.  This procedure was successful at eliminating objects 
completely except when fewer than 10 images were available to form the median.
To
eliminate artifacts from such residuals, we smoothed the skyflat using a 
variably sized boxcar filter via the IRAF {\sc mkskycor} task.
Each skyflat was then normalized by the mode of its central area
and divided into the images, resulting in a reduction of the
gradients and variations across each image to at most the 2\% level.  
To eliminate these remaining random variations, which could be found in 
every filter except U, we used 
the Faint Object Classification and Analysis System ({\sc focas}, \cite{val82a})
sky-fitting routine to estimate the local sky value for every 8x8 pixel box. 
The sky values determined by {\sc focas} were found to be accurate to within 1\% even 
in areas dominated by saturated stars, but were offset from their true 
locations slightly due to the line-by-line process of sky determination 
used by {\sc focas}.  To avoid creating troughs to one side of bright objects after
sky subtraction, we produced three additional {\sc focas} sky images, rotating the 
image by 90\arcdeg\ between each.  These sky images were rotated back to the 
original orientation and medianed with the first sky image, discarding 
the highest pixel, to form the final sky image.  Each field image had its own 
sky image subtracted from it, leaving fields globally flat to $<$0.5\%.
The only shortcoming of this method was that the outer regions of the largest, 
most extended galaxies present (typically 20--50\arcsec\ in size) were sometimes
confused with the sky, resulting in an oversubtraction of the sky around these
galaxies.  This effect is not of concern for the catalog of stellar objects.

Next, cosmic rays were isolated and interpolated across using the IRAF 
task {\sc cosmicrays}.
The removal worked quite well,
having trouble only with some many-pixel cosmic ray hits and almost 
never flagging unsaturated objects by mistake.  To allow us 
to be conservative in our eventual selection of candidates in the event that 
the interpolation introduced errors into the object photometry, 
all cosmic ray locations were added to the bad pixel lists for later 
cross-correlation with the object lists.

Lastly, the images were shifted and trimmed so that each CCD field was on a 
common coordinate system to the nearest pixel (no fractional pixel shifts
were used).  The area covered by these trimmed images, taking into account the
additional reduction from adjacent field overlaps and the 5\arcsec\ wide border
on each image unused by {\sc focas}, is given in Table 1.  The total areal 
coverage on the sky of the survey is 2990\sq\arcmin, or 0.83\sq\arcdeg.  

\section{Detection and Cataloguing of Objects}

Because of the variations in seeing, focus, and pointing between the two images
in each filter in our fields (which were often taken on different nights), as 
well as the problem of cosmic-ray contamination, we decided not to coadd the 
images to produce a single image with improved signal-to-noise (S/N) in each 
filter.  Instead, we reduced each frame individually and improved the
S/N by averaging the magnitudes from each frame after they had been reduced.
The main drawback to this approach with respect to coadding is the increased 
unreliability of object classification at faint magnitudes, but we are helped 
in this by having two independent classifications.
We used {\sc focas} and its `built-in' detection filter to detect and classify all 
the objects in our frames, and the IRAF {\sc phot} routine to find aperture 
magnitudes for them.  

A brief summary of the detection and initial photometry steps performed on 
each field follows: 

\noindent{$\bullet$ {\sc focas} detections must have at least 9 contiguous pixels 2.5$\sigma$\ above the sky, after the value of each pixel was convolved
with the values of pixels within a 5$\times$5 grid, weighted according to
the standard `built-in' {\sc focas} detection filter}

\noindent{$\bullet$ Objects lying on bad pixels were flagged as such}

\noindent{$\bullet$ Objects were classified using a different empirical PSF for 
each of nine subsections on each image}

\noindent{$\bullet$ Objects not appearing on both images in the filter were discarded}

\noindent{$\bullet$ To be considered stellar in appearance, objects were 
required to be classified as a star in at least one image of half or more 
of the filters in which they were detected}

\noindent{$\bullet$ Aperture corrections were calculated for each image 
subsection}

\noindent{$\bullet$ Instrumental magnitudes were calculated in a 3\farcs0\ 
diameter aperture}

\noindent{$\bullet$ The instrumental magnitudes were calibrated to a standard 
system}

\noindent{$\bullet$ Magnitudes from each image in each filter were combined, 
accounting for bad pixel contamination and magnitude limits, to give a final 
magnitude for each object.}

Below, we discuss important reduction steps in more detail, where necessary.

\subsection{Bad Pixels}
All objects that fell on bad pixels were flagged as such.  `Bad pixel' here is 
a broad term, including sites of detected cosmic ray hits, 
the actual bad columns and traps determined through inspection, and the bad 
focus regions (typically the corners) of each image.  
Saturated stars were also
considered to fall on bad pixel areas, but were flagged differently since the
determination of color limits for such objects is different than for regular
bad pixels (we measure a lower limit to the object's magnitude in these cases).

\subsection{Classification}

Objects were classified using the standard {\sc focas} method and parameters,
using the resolution classifier fully described in Valdes (1982b) and
summarized briefly here. An empirical point-spread function (PSF) is 
constructed from isolated bright stars.  From the PSF a set of 
classification templates is generated,
consisting basically of a broadened or narrowed PSF.
Objects were then classified as star, fuzzy star, galaxy, diffuse object, or 
noise spike depending on which template they best fitted.  
Due to the frequent variation in focus across the chip, each image was 
divided into 9 subsections in a 3x3 grid, and a different empirical PSF 
was used for classification in each subsection.  This method is obviously 
not as accurate as fitting a slowly varying PSF across the chip; but the 
latter method is difficult to implement with {\sc focas}.  
We included `fuzzy stars' in our stellar object list to increase our 
ability to include objects where an active nucleus is seen along with extended 
light from a host, companion, or intervening galaxy.  Inspection of the 
location of objects on a plot of PSF scale vs. nuclear component fraction 
showed that only $\sim$10\% of fuzzy stars have an extended component with
amplitude significantly different from zero.
This tells us that extended objects with strong 
nuclear light components are relatively rare, and reassures us that we are
not including a large number of galaxies with this criterion. 

It should be noted that some of the frames had a strong gradient in focus 
across them, causing substantially out of focus images at the corners of the 
chip.  Objects in these regions were flagged so that the photometry from them 
is not used; fortunately, no region was out of focus in both offset images of 
the field. 

\subsection{The Stellar Object Catalog: Selection, Completeness, and Misclassification}

Both offset frames in each filter were matched with each
other, and objects not detected in both offset images were discarded.  An 
object was classified stellar {\it in a given filter} if it was classified by 
{\sc focas} as a star or fuzzy star in {\it either} offset frame of the filter.
The criteria for inclusion in the final catalog of stellar objects
were that the object be detected in at least three filters and 
classified as stellar in half or more of the filters in which it was detected.
Our criteria are rather lenient in allowing objects classified as nonstellar in 
some images to be included in the stellar object catalog because we did not
want to exclude faint AGN with detectable host galaxies.  We need to understand
how often nonstellar objects are misclassified as stellar, and vice versa, as a 
function of magnitude in each filter in order to understand the 
contamination and completeness of our catalog.

The ideal method of determining the completeness and classification reliability
is to add to the original images simulated objects of known morphology and a 
wide range of magnitude.  These images are then processed as the original 
images were, and the completeness and classification reliability are 
straightforward to determine.
However, the varying PSF over many of our CCD images makes the addition of 
accurately simulated objects to the data complicated.  Instead, we opted to 
create entirely artificial images and analyze them.  Since the parameters of 
the artificial data can be made to match those of the real data very 
accurately, this approach should yield an accurate description of the real 
data's classification reliability and completeness as a function of magnitude.

We used the IRAF {\sc artdata} package to simulate two images of a CCD field
in each filter.  We attempted to match the seeing, background, and noise level
observed in the 01w field.  
For the galaxies, to subjectively match the appearance of the real data, 
a characteristic radius of 5.3'' 
and a mixture of 40\% de Vaucouleurs profile `bulges' and 60\% exponential
profile `disks' was chosen.
Galaxies were given a power-law magnitude distribution with slope 0.35 
(i.e. n(m)$\sim$10$^{0.35m}$), which matches well the slope of faint magnitude 
galaxy counts.
To increase the number of faint stars available for studying completeness,
stars were given a power-law magnitude distribution with slope 0.6, steeper 
than the observed distribution.
However, since this distribution does not generate as many bright stars as the 
real images contain, it was necessary to add a grid of bright stars to the 
images to ensure that the PSF could be well determined in each image subsection.
Since whether or not an object is included in our catalog of stellar objects
depends on its classification and magnitude in several filters, objects 
of different colors had to be simulated.  
We divided the locus of stellar objects in our real data into five bins, 
each with different broad-band colors.
The relative numbers of stars and galaxies in each bin was determined from the 
real data, and corresponding percentages of the simulated stars and galaxies 
were assigned the colors of each bin.
Finally, following the above parameters, 3000 stars and 2500 galaxies with 
R=16-24 were randomly distributed across the field.  
The appearance and statistics of the images closely matched those of the real
data.  One minor difference is that the PSF in the real data is not a perfect 
gaussian.  However, the difference between the real PSF and the simulated one 
is only apparent in the wings of bright stellar objects, and so this difference
should not affect any of our results on faint objects.

Our detection and cataloging programs were then run on the artificial images 
in exactly the same manner as on the real data, producing an output {\sc focas}
catalog of stellar objects.
The {\sc focas} task {\sc artcat} was used to create {\sc focas} catalogs
of the stars and galaxies input to the {\sc artdata} task, for comparison
with the output catalog.

To determine its completeness, the output catalog of stellar objects was 
compared to the input catalog of stars.
As is typical for CCD data (\cite{ste90}), the completeness fraction is 100\% 
at bright magnitudes, begins a slow linear decline at some faint threshold
magnitude as the effects of crowding appear, and then switches to a rapid linear
decline as the detection threshold comes to dominate over crowding effects.
We find that this transition occurs at about the 90\% completeness level, which
closely approximates the 5$\sigma$\ detection limit (see \S 4.5).
Similarly, the 50\% completeness limit closely approximates the 
3$\sigma$\ detection limit.
These limits are listed in Table 3, along with the threshold magnitude
above which the completeness is essentially 100\%.

To determine the classification reliability, the output catalog of stellar
objects was compared to the input catalog of galaxies to determine which
supposedly stellar objects were actually misclassified galaxies.
In principle, false detections due to noise could also show up in the catalog,
but by requiring detection in both offset frames of at least three filters we
completely eliminate such detections.
376 galaxies were misclassified as stars in our simulations, virtually none of
them brighter than the completeness threshold magnitude, which we thus adopt as
the threshhold magnitude for classification as well.
To apply these results to the real dataset requires adjusting for the expected 
number of faint 
galaxies in the survey fields versus the number used in the simulated data,
since the number of misclassified galaxies in the catalog obviously depends
on the number of galaxies that exist in our fields.
We used the compilation of McLeod \& Rieke (1995) to determine the number of
faint galaxies expected in our fields as a function of magnitude.
The slope of the N(m) relation for our simulated galaxies is 0.35, close
enough to the value observed for faint optical counts that we only need to 
adjust the normalization.
To approximate the counts from R=21.5--23.5 we used the counts from Gunn 
r=22--24 and assumed that those counts are 100\% complete.
In our 0.0713\sq\arcdeg\ simulated field, the input catalog contained 1761 
galaxies from R=21.5--23.5, whereas $\sim$50\% more (2797) are expected in an 
equal area of real data.
Thus to make our simulated field accurately reflect the number density of faint
galaxies, we adjust the total number of contaminating galaxies upwards by 50\%, 
from 376 to 564.
In our entire survey area of 0.83\sq\arcdeg, we then expect 6565 spurious 
objects in the stellar object catalog.

\begin{figure}[t]
\plotone{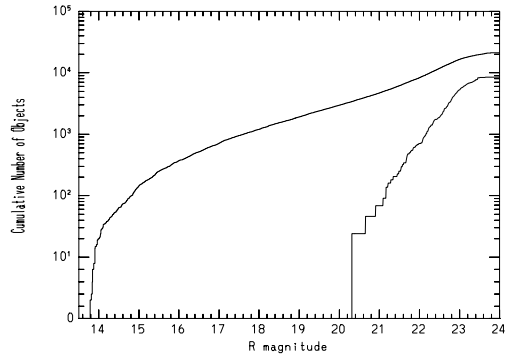}
\caption{
The upper curve is the observed cumulative number-magnitude (R band) counts of 
all 21375 objects in the stellar object catalog.
The lower curve is the expected cumulative number-magnitude (R band) 
distribution of misclassified galaxies in the catalog, based on scaling 
the results of the artificial data simulations described in the text to 
the expected galaxy number density and total area of the survey.
At faint magnitudes (R$\gtrsim$21.5) the stellar object catalog shows an
excess above the power-law distribution seen at intermediate magnitudes,
which may reasonably be attributed to misclassified galaxies.  
}
\end{figure}

Using this scaling and the magnitude distribution of the misclassified galaxies
in our simulation, in Figure 2 we compare (in the R band) the expected number 
of contaminating galaxies to the total number-magnitude counts of the stellar 
object catalog.
At faint magnitudes (R$\gtrsim$21.5) the stellar object catalog shows an 
excess above the power-law distribution seen at intermediate magnitudes.  
This may reasonably be attributed to misclassified galaxies.  A further 30\% 
increase in the number of contaminating galaxies, to 8535, is required
to provide a good match to the number-magnitude distribution of this excess.

What might have caused our simulations to underpredict the number of
misclassified galaxies?  There are some effects we have neglected in the 
simulations, but most of them should not alter our conclusions significantly:

\noindent{
$\bullet$~We have not included cosmic rays in the artificial data, although 
this capability does exist within {\sc artdata}.
Cosmic rays would effectively act as another source of noise, and since our 
selection criteria are very efficient at keeping false detections due to noise 
out of the catalog, their effect is negligible.
}

\noindent{
$\bullet$~We have not included the effects of interpolating over bad pixels, 
although this is not difficult to do either.  Interpolation might cause objects
to fall below the detection threshold or be misclassified.  Since
the area affected by bad pixels is very small and its location varies from 
image to image, and since our criteria for inclusion in the final catalog uses 
information from many images, the effect of bad pixels should be negligible.
}

\noindent{
$\bullet$~We have ignored the effects of galaxy clustering, since its amplitude
is very weak at these magnitudes (\cite{nw95,roc93}).
}

\noindent{
$\bullet$~In reality, different fields have different values for the seeing, 
galaxy and star number counts, etc., than those used in the simulated data.  
However, the 01w field that we simulated is fairly typical in these respects, 
so generalizing the results of the simulated 01w data to the entire survey 
should be valid.
}

\noindent{
$\bullet$~We have not taken into account that one of our fields, the 01e 
field, has only one offset image in each filter, so that its completeness and 
contamination properties will be different than those of the other fields.  
This will likely raise the contamination fraction and possibly reduce the 
completeness slightly, but since the area affected is $<$10\% of the total
the effect should be negligible.
}

\noindent{
$\bullet$~One possibly significant effect not included is the variable PSF 
over many of our images, due mostly to a softening of the focus at increasing 
distances from the center of the field.  This will in general have the effect
of increasing the contamination, since the classification of faint objects is 
more difficult when the PSF is broader.  However, only a small fraction of the 
total area of the survey has an extremely broad PSF, so its effects should not
greatly alter the conclusions of our simulations.
}

The most important potential problem with our simulations is the scale radii
used for the simulated galaxies in them.
Recent results from the $HST$ Medium Deep Survey (MDS; \cite{mds94,im95})
suggest that many faint galaxies have very small half-light radii
(median 0\farcs4 at V=21--24).  
Naturally, galaxies with small scale radii are more easily misclassified as 
stars than galaxies with large scale radii under any seeing conditions.
Our simulated galaxies have half-light radii from 0\farcs5--1\farcs5 at faint
magnitudes; thus, we may have underestimated the contamination fraction in the
stellar object catalog.  Based on the fraction of our simulated galaxies 
misclassified as stars as a function of their scale radii, we estimate that 
using the MDS scale radii distribution would lead to a doubling of the 
contamination fraction in the stellar object catalog.
The contamination fraction is also dependent, albeit to a lesser degree, upon 
the relative numbers of galaxies with de Vaucouleurs and exponential profiles, 
and the inclination angle distribution of the latter population.  However, 
these are second-order effects, and our chosen values for the parameters 
involved should be valid for all but the faintest magnitudes (cf. \cite{im95}).

The MDS results are preliminary, being based on a small number of galaxies;
also, due to the small area studied so far, there is very little data on the 
distribution of galaxy scale radii at brighter magnitudes.
It is thus perhaps not surprising that results from the Canada-France Redshift 
Survey show that only 1\% of all galaxies at 17.5$<$I$_{AB}$$<$22.5 
($\sim$19.5$<$B$<$24.5) are as compact as stars on images with 0\farcs9 
seeing (\cite{cf5}).
While our seeing is rarely that good, this result indicates that the
distribution of galaxy scale radii used in the simulations may not be as 
unrealistic as the MDS results suggest, particularly at brighter magnitudes.
In the next few years more extensive data from the MDS should
yield very accurate distributions of galaxy scale radii, morphological types,
and axis ratios over a large magnitude range.
This will enable more accurate simulations to be done to help understand the
results of ground-based imaging.
Based on the preliminary MDS results, we feel that an upwards adjustment of 
30\% in the contamination fraction of the stellar object catalog is reasonable.
(Recall that this adjustment results in a excellent fit to the observed excess 
of faint stellar objects above an extrapolation of the counts at brighter
magnitudes.)

Thus, in summary, we adopt as the total contamination fraction of our catalog 
the prediction of our simulated data, scaled to the expected number density of
faint galaxies and the size of our survey, then adjusted upwards by 30\%.
Since the catalog contains 21375 objects, its overall contamination fraction
is 40\% (8535 objects), with roughly half of that occurring fainter than the 
5$\sigma$\ limit.
We parametrize the integral contamination in each filter (i.e. the total number
of misclassified galaxies in the catalog brighter than a given magnitude in 
that filter) as a cubic function rising from zero at the threshold magnitude 
to its maximum value at the 3$\sigma$\ limiting magnitude.  
These magnitudes are tabulated for each filter in Table 3, along with the
integral contamination at the 5$\sigma$\ limiting magnitude.
The completeness of the catalog in each filter is 100\% down to the threshold 
magnitude and then declines linearly to 90\% at the 5$\sigma$\ limiting 
magnitude and 50\% at the 3$\sigma$\ limiting magnitude.
The contamination ($\sim$25\%) and completeness (90\%) of the stellar object 
catalog at the 5$\sigma$\ limits are well characterized and allow for reliable
corrections to be made when computing number counts, etc.

\subsection{Photometry and Aperture Corrections}
Aperture photometry of all stellar objects was performed with the IRAF 
{\sc phot} task, using a centroiding algorithm to determine the position 
of the aperture center.
The instrumental magnitudes were calculated using a 3-pixel (1\farcs6) radius 
circular aperture and a sky annulus 2\farcs7 wide 8\arcsec\ away.
This was found to be the best compromise between maximizing the enclosed 
signal and reducing the scatter between different measurements of the same 
object due to focus and seeing variations between different exposures.
Aperture photometry was performed at the object's estimated coordinates
in all filters, even if the object had not been originally detected in
the particular filter, in order to obtain color limits on such objects.
In such cases, occasional centering errors would occur and need to be 
corrected --- the centering routine would latch on to a nearby object
or noise spike and center the aperture in that position, rather than the
object's expected position.  

Aperture corrections were used to correct the magnitude measured in a 1\farcs6
aperture to the 5\farcs8 aperture used for the standard stars.
The corrections were interactively calculated for the 9 subsections on each 
image separately, using up to 20 of the brightest stars in each subsection.
The final variations of the aperture corrections across the chip were found
to be acceptably small and to yield good independent magnitude measurements 
from the two offset frames.  

\subsection{Final Magnitude Determination}

Before the determination of final magnitudes, the instrumental magnitudes 
were converted to a standard system, as detailed in \S 5.
Once the calibrated magnitudes were available for each offset image,
the final magnitude for each object was determined as follows:

\noindent{$\bullet$ use the average of the two measurements, unless one
measurement lies on a bad pixel, is saturated, or is a limit}

\noindent{$\bullet$ if the object lies on a bad pixel or is saturated in 
both images, use the average, but flag it as possibly being in error}

\noindent{$\bullet$ if measurement A is a limit, and B is a detection below 
frame A's limiting magnitude, take the detected value from B as the magnitude}

\noindent{$\bullet$ if measurement A is a limit, and B is a detection above 
frame A's limiting magnitude, average the two and set the error negative 
(this last step is done to help mitigate the effects of spurious detections 
caused by overlooked bad pixels)}

\noindent{$\bullet$ if the final error from the previous steps is greater 
than 0.333, i.e. the object is fainter than our 3$\sigma$\ detection limit, 
replace the magnitude of the object with the 
3$\sigma$\ magnitude limit of the combined frames, and the error with 0.333}

The {\it average} 3$\sigma$\ magnitude limits in our survey are given for 
each filter in Table 3.  Technically, 3$\sigma$\ detections only require 
${\sigma}<$0.333, regardless of the object's magnitude, but it is useful 
to know the average magnitudes to which this limit corresponds.
Also, upper magnitude limits for each filter, established by examining the 
magnitudes of saturated objects, are given in Table 3.
Only a handful of saturated objects were found fainter than those limits.

\section{Photometric Calibration}

Standard stars in M67, M92, and NGC7790
were observed on all nights of both observing runs.
Standard magnitudes and color indices were taken variously from Johnson and 
Sandage (1955), Eggen and Sandage (1964), Sandage (1966), Schild (1983), 
Christian et al. (1985), and Odewahn et al. (1992).  These data, supplemented
with unpublished photometry from A. Porter and L. Davis, are available through 
the IRAF {\sc photcal} on-line library.  

Standard star images were reduced in the normal manner for CCD images, using 
the same domeflats and skyflats used for the field images.
Instrumental magnitudes were found for all standard stars using 5\farcs8
radius circular apertures, with all other parameters, including the sky 
annulus size, being the same as used for the field images.

The photometric solution for the \ubvr~filters is independent of that for our 
two I filters, and is discussed first.  In the former case we transform 
our observed magnitudes to the standard \ubvr~system; in the latter we
define a new system for the I filters.

\subsection{\ubvr\ Calibration}

The IRAF {\sc photcal} photometric calibration package was used to 
interactively solve
the following system of equations involving the instrumental magnitudes 
(lower case), the standard calibrated magnitudes and colors (upper case), and 
the airmass for the observations in each filter (X$_u$, X$_b$, X$_v$, X$_r$).  
The system of equations was solved simultaneously for the zeropoints 
(u$_0$, b$_0$, v$_0$, r$_0$), the color coefficients 
(u$_1$, b$_1$, v$_1$, r$_1$), and the extinction coefficients 
(u$_2$, b$_2$, v$_2$, r$_2$) for each night separately.

\begin{equation}
u = u_0 + u_1 \times (\ub) + u_2 \times X_u + V + (\bv) + (\ub)
\end{equation}
\begin{equation}
b = b_0 + b_1 \times (\bv) + b_2 \times X_b + V + (\bv)
\end{equation}
\begin{equation}
v = v_0 + v_1 \times (\vr) + v_2 \times X_v + V
\end{equation}
\begin{equation}
r = r_0 + r_1 \times (\vr) + r_2 \times X_r + V - (\vr)
\end{equation}

Not all nights yielded well-defined photometric solutions independently, since
typically only three observations of standard star clusters were made each
night.  Since most parameters in the solutions had the same values within the
errors from night to night, and indeed between the two runs, we decided to 
adopt the same color terms and extinction coefficients for each observing run
and let only the zeropoint in each filter vary.  Thus we solved the equations 
again, this time only for the zeropoint, keeping the other variables fixed at
the average values determined from the individual nights' data.  The resulting 
final \ubvr~photometric calibration parameters for both runs are listed in 
Table 4.

As shown in Table 4, the formal errors associated with the photometric
parameters are a few percent at most.  However, differences from the base 
zeropoints were found to exist for most CCD images of the survey fields in 
both runs, as discussed in \S 5.3.

\subsection{I75 and I86 Calibration}

Our magnitude system in the I75 and I86 filters is {\it defined} by 
observations of the spectrophotometric standard DA white dwarf Wolf 1346 
made on 1990 August 24. The definition and calibration of our system 
to the AB system are 
described in detail in this section.

Instrumental magnitudes for Wolf 1346 were determined using a 
(8\farcs5) radius circular aperture.  AB magnitudes for the star were taken 
from the IRAF {\sc onedstds} on-line library, 
which in the case of Wolf 1346 consists of unpublished data plus data from 
Oke (1974) for wavelengths redward of 8400~\AA.  Despite its greater noise, 
we chose to use the Oke data rather than the data from Massey et al. (1988) 
because of its better coverage of the 7000-9500~\AA~wavelength region.

In order to find the effective wavelengths of the filters for this star, 
the AB magnitudes were converted to fluxes
and were convolved with the filter + CCD + telescope system response curves.  
The effective wavelengths for Wolf 1346 in the I75 and I86 filters are 
7430~\AA~and 8580~\AA and the
standard magnitudes of Wolf 1346 are thus defined to be its interpolated
AB magnitudes at these wavelengths: I75=12.02 and I86=12.22.

The extinction coefficients at the wavelengths of these filters was 
calculated in two ways: first by using the standard KPNO extinction curve from 
the IRAF {\sc onedstds} library to find the `effective extinction coefficient' 
and second by 
comparing measurements of the instrumental magnitudes of stars in NGC 7790 
made at different airmasses during the same night of the 1990 August run.  The 
extinction coefficients for I75 found by the standard and empirical methods 
were 0.063 and 0.078$\pm$0.014 respectively; and for I86 they were 0.039 and 
0.039$\pm$0.008.  We chose to adopt the empirical values, but the difference
between using the different values would be
only a few percent at most, since 94\% of our observations were made 
at airmasses less than 1.4 and 100\% at airmasses less than 1.65.

The final step in determining the zeropoints is simply one of plugging in to
the equation
\begin{equation}
m_{instr}=m_{AB}+zeropoint+X*airmass
\end{equation}
which {\it defines} the zeropoints of the I75 and I86 filters for the 1990 
August 
run.  These zeropoints were found to be -22.08 and -21.53, as given in Table 4.
Note that no color term was adopted for our two I filters.

In order to obtain the I75 and I86 filter zeropoints for the 1991 April 
run, the standard stars in the standard fields observed during both runs (M92 
and NGC7790) were defined as secondary standards in the I75 and I86 filters as 
follows.  Using the data from 1990 August, equation (5) 
was solved for m$_{AB}$ for each standard star in each cluster, holding fixed 
the zeropoints and the extinction coefficients at the values given above.  
The values of m$_{AB}$ derived from these 1990 August observations were
adopted as the magnitudes for these standard stars in our two I filters.
The zeropoints for the 1991 April runs were then found by solving the standard 
photometric transformation equations in the same manner as equations (1-4)
for the \ubvr~filters, with the exception that no color terms were used.  
Once again, the extinction coefficients were held 
fixed, because we assumed that the extinction did not change markedly between 
the two runs, which was certainly the case with the \ubvr~data.

\begin{equation}
i75 = I75 + i75_0 + i75_2 \times X_{i75}
\end{equation}
\begin{equation}
i86 = I86 + i86_0 + i86_2 \times X_{i86}
\end{equation}

These zeropoints were found to be -22.03 and -21.48, as given in Table 4.  
These values are only 5\% different from those for 1990 August.
Again, as for the \ubvr~filters, some differences from the base zeropoints in 
were found to exist for most CCD images of the survey fields in both the 
1990 August and 1991 April runs, as listed in Table 2.

\subsection{Further Empirical Calibration}

The photometric calibration derived from the standard stars was checked on the
survey fields by calibrating the instrumental magnitudes using only the 
zeropoint and airmass coefficient at first, i.e. without including the color
term.  This allowed us to check for any zeropoint shifts between CCD images
of the same field and make adjustments to the zeropoints as follows.  
We plotted the difference between the magnitudes measured for all 
`good' objects (ones not affected by bad pixels) in the two 
independent offset frames of each field versus the average of the two 
magnitudes (hereafter we refer to these as the `magnitude difference' graphs).
A typical magnitude difference graph is shown in Figure 3.
Figure 3 shows that at bright magnitudes the two measurements agree to 
within the small ($<$2\%) uncertainty due to limitations of the 
flatfielding, etc.  It also shows that our bad pixel removal is quite good, 
with only 2 or 3 objects (out of $\sim$1000) showing differences between the 
two magnitude measurements large enough to be undisputably due to errors.
If the average magnitude difference in a plot like Figure 3 is not zero, one of
the frames has been calibrated with an erroneous zeropoint.  
The magnitude difference graphs are only useful in calculating zeropoint shifts
when the same set of objects has been measured; so, in order to measure 
zeropoint shifts between frames in different fields, we plotted and compared 
various color-color diagrams as detailed below.

\begin{figure}[t]
\plotone{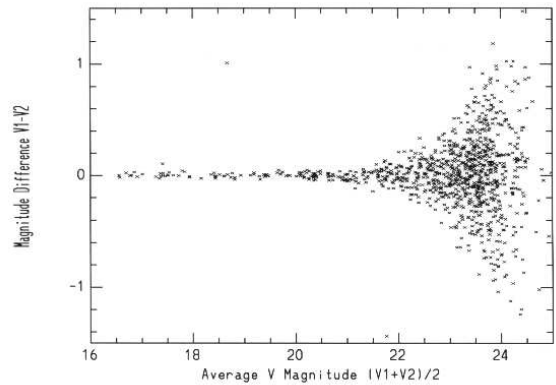}
\caption{
A typical magnitude difference graph, in this case from the V images
in the 01w field.  The x-axis is the average V magnitude measured in the two
images for each object, and the y-axis is the difference between the magnitudes
measured in the two frames for all `good' objects (ones not affected by bad
pixels or focus problems).
}
\end{figure}

It became apparent from these graphs that empirical corrections to the 
photometric solutions in most of our fields were necessary in order to 
bring all fields onto a common system, similar to the practice in 
astrophotography of calibrating each photographic plate independently.  
There are many possible explanations for zeropoint differences between fields 
-- the possible presence of light cirrus as noted in our observing logs, as 
well as night-to-night variations in the atmospheric transparency and 
extinction, being the most prominent -- but for our purposes the exact causes 
are not relevant.  
This is because one of the main reasons we calibrate the photometry is to 
bring all the fields to a common system where the locus of stellar colors shows
as little scatter as possible, which helps us better discriminate candidate 
quasars from stars.  
It is not as efficient to search for outliers in each field independently
because of the relatively small number of objects that define the stellar locus
in each field.  That is, a slightly unusual but normal star (e.g. O, B, and A 
stars at high galactic latitude) will stand out more in the color-color 
diagrams of each individual field than in the combined diagram of all fields.
Strictly speaking, our empirical corrections should be made to the extinction
terms in the photometric solution, since they are the most likely to vary. 
However, it was easier to derive a correction to the zeropoint of each
field rather than to the extinction term, so this was done instead.
Since the shift needed to align the stellar locus of a given field is fixed,
the term used to make that adjustment matters little in practical terms.

The 17n field was adopted as our best field, relative to which all others
were adjusted.  It showed an average magnitude difference between offset frames
of zero in all but the U and B filters, and in those filters it was obvious by
inspection and by consulting our observing logs which two frames were 
affected by light cirrus.  Thus we believe that the 17n magnitude calibration
is exact to within a few percent.  Note, however, that since we have tied our 
magnitude calibration to this field, any systematic error in the 
calibration assumed for it will affect all our other fields as well.  
Since the 17n field was observed in the 1991 April run, all other 1991 April 
fields had their zeropoints adjusted relative to the 17n field by direct
comparison of several color-color diagrams.  For both sets of offset frames
in each field, color-color diagrams were plotted on an identical scale using 
the brightest one hundred unsaturated non-bad-pixel objects.  Each field's 
diagrams were compared with those for the 17n field and were shifted by eye 
in both colors until the best alignment was obtained.  
The empirically required shift in a given color is simply a combination of the
empirical zeropoint shifts required for the two filters comprising the color.
By combining the shifts required for five (independent) colors with the shifts 
required between frames of the same filter in each field, we are able to 
determine each field's zeropoint shifts relative to the 17n field 
to within an additive constant which is the same for all filters.  That is,
we can determine the relative zeropoint shifts between all filters, but 
there is still an overall additive constant for the zeropoint in all filters 
that we cannot determine in the same manner.
This additive constant was determined by requiring that all empirical R-band 
zeropoint shifts be as small as possible relative to the 17n field.  
This is probably a better estimate than requiring the average zeropoint shift 
in all filters to be minimized, because the U, B, and V filters will be 
affected by varying atmospheric extinction more than the R band.
For our 1990 August run, we used the fact that the 17s field, which was 
observed at that time, overlaps with the 17n field.  Approximately twenty of 
the brightest (R~$<$20) objects in the overlap region were used to find the 
exact shifts between the zeropoints for the two fields by requiring that the 
average
difference between these two independent measures of the magnitudes of these
objects be equal to zero.  The zeropoints of 17s were adjusted as needed to
achieve this, and then the other 1990 August fields were adjusted relative to 
the 17s field in the same manner as the 1991 April fields were adjusted 
relative to 
the 17n field.  The average shifts in each filter were less than 5\% half the
time, and less than 10\% in all cases except the R filter in the 1990 August
run, where the empirically determined shift was 0.13 magnitudes, or 13\%,
different from its value calculated by the photometric comparison.
While this is uncomfortably high, comparison of color-color plots from 
different fields shows clearly that our empirical zeropoint shifts achieve
the desired result of tightening the stellar locus.
Since the empirical shifts in the zeropoints are typically between 5~and~10\%,
we estimate that the systematic uncertainties in our magnitude calibration
are at this level.

As a final check, we used bright (R$<$~20) objects, typically two dozen of
them, in the overlap regions between the 21-hour and 22-hour fields to 
confirm our determinations of the zeropoint shifts for the different filters
and fields.  The magnitudes agreed to within 2$\sigma$\ in all cases.  
The final zeropoints for each CCD field in both runs are given in Table 2.
Note that for all filters in the 01e field and the B filter in the 10e field, 
only 1 image was obtained.  Thus no magnitude difference plot could be 
constructed, and calibration of the zero points was performed only by 
color-color diagram comparison.  This should not introduce any significant 
additional uncertainty into the calibration of these fields.

\subsection{Photometric Error Analysis And Discussion}

We also used the magnitude difference graphs to study the accuracy of 
the photometric errors determined by the IRAF photometry routines, to ensure 
that there were no unexpected additional sources of error that had 
not been taken into account.  

The magnitude difference is d$\equiv$ m$_1$-m$_2$, where m$_1$$\pm\sigma_1$ and 
m$_2$$\pm\sigma_2$ are values of the magnitude of the same object measured from 
independent CCD frames.  Thus the distribution of magnitude differences at 
any particular magnitude should be given by a gaussian:
\begin{equation}
f(d(m_{avg}))=(1/\sigma_d\sqrt{2\pi})e^{-[d(m)]^2/2\sigma_d^2}
\end{equation}
where the magnitude m$_{avg}$=(m$_1$+m$_2$)/2 and the sigma of this distribution is
$\sigma_d^2(m_{avg})=\sigma_1^2+\sigma_2^2$.
It is reasonable in most cases to assume $\sigma_1=\sigma_2$, i.e. that 
the photometric errors in each CCD frame of a particular filter and field are 
the same as a function of magnitude.  
Thus if we measure $\sigma_d$, we can find $\sigma$ from the equation 
${\sigma}=\sigma_d/\sqrt2$.

We obtained two different estimates of $\sigma_d$ from our magnitude difference
plots.  First, we took the square of the magnitude difference for every object
not affected by a bad pixel, 
sorted by magnitude, binned them in groups of 50, and calculated the 
average value of the square of the magnitude difference for each bin, 
Second, we took the absolute value of the magnitude difference for each object,
sorted and binned them as before, and calculated the average absolute value
in each bin.  This gives us a measure of the mean deviation 
$\alpha_d=\int|d|f(d)$=0.8$\sigma_d$ (\cite{you62}).
 
In Figure 4 we plot the photometric errors from the IRAF {\sc phot} routine for 
all objects in both CCD frames in a representative field and filter and the 
sigma estimated from the magnitude difference plots in both manners described
above.  The match is clearly quite good.
At the brightest magnitudes we have replaced the photometric errors determined 
by IRAF with 0.02 if they were originally smaller than this (to be exact, this
error should be added in quadrature to all the objects' errors).

In summary, the external errors in our photometry, as estimated from the two 
independent measurements of the magnitudes of each of our objects, have been 
shown to be consistent with the internal errors computed according to photon 
statistics, except for a $\sim$2\% additional uncertainty independent of
magnitude.  
This independent check proves that our flatfielding process, aperture 
correction procedures, and photometry methods are all quite reliable, 
having inherent limitations of only the aforementioned $\sim$2\%.
As for systematic errors, our stellar locus matches values for stellar colors 
from the literature to about 5\%.
An exact match with expected colors is unimportant in terms of outlier
selection, but necessary for accurate comparison with other studies and for
other uses of the dataset.

\section{Astrometry}

The final step in the preparation of the catalog was the conversion of CCD 
(x,y) coordinates (centroids) to right ascension and declination.
We required accuracy to $<$0\farcs5
because the optical fibers used by the HYDRA instrument for our initial
spectroscopy are only 2\farcs0 in diameter.  The astrometry was performed
by A. Porter in the manner briefly outlined here.

The program {\sc FINDER} (\cite{and92}) was run on each survey field to search
for HST Guide Star Catalog (GSC; \cite{taf90}) stars within the field. 
There were sometimes less than a dozen or so GSC stars in a field.
In these cases we used POSS plates to construct an astrometric grid
based on nearby AGK stars.
The positions of all the GSC stars in the field, plus a dozen or two
anonymous field stars, were then measured on this grid.
This served as a check on the GSC astrometry (which was found to be accurate to
$<$0\farcs5 in our fields, adequate for our purposes), and yielded enough stars
(between 14-26 for each CCD field) to construct a good astrometric solution.
The CCD (x,y) centroids for all these astrometric calibration stars were 
measured on the first R frames of each field using the IRAF task \sc imexamine.
\rm The astrometric solutions for CCD (x,y) to (RA,dec) conversion in each field
were calculated using the program {\sc ASTRO} (\cite{and92}) and included 
terms up through a r$^2$ barrel distortion term.  
The internal residuals of the solutions ranged from 0\farcs18 to 0\farcs46, 
and no significant trends or distortions in the solutions were present.
The average positional agreement between stars in the overlap regions of 
adjacent
CCD fields was usually better than 0\farcs1 and always better than 0\farcs4.
The solutions were then used to find the RA and Dec of all objects in the 
catalog, using the CCD (x,y) positions of the objects as measured from the 
first R frame of each field.

\section{Summary}

The final photometrically calibrated catalog of stellar objects in our Deep
Multicolor Survey contains 21,375 objects from 0.83\sq\arcdeg\ of sky.  
All objects have positions accurate to $\sim$0\farcs5 in each coordinate and 
magnitudes in 
six bands whose errors are consistent with expectations from photon statistics.
The production of the catalog is the main result of this paper.
From this catalog we would like to isolate as many as possible of the $\sim$193
quasars with B$\leq$22.6 expected in our survey area (\cite{maj93}),
with a minimum of contamination from stars and other objects.
As there are 4,205 objects with B$\leq$22.6, quasars are outnumbered by a 
factor of 20:1, and careful candidate selection is obviously necessary.
Paper II (\cite{dms2}) discusses our candidate selection methods and presents
initial spectroscopic results and conclusions.

%
%
%
%
%
%

\acknowledgements

We thank Jeannette Barnes for invaluable assistance (and endless patience) 
with our IRAF questions, Frank Valdes for essential help with {\sc focas}, 
Todd Boroson for providing us with his `superflat' flattening routine, Lindsey 
Davis for providing us with some unpublished standard-star photometry, the KPNO
mountain staff, particularly the telescope operators, for their assistance in 
observing, and last but definitely not least the KPNO TAC for their allocation 
of time for this project.
PBH acknowledges support from an NSF Graduate Fellowship and a University of
Arizona Graduate College Fellowship.

\clearpage


\begin{deluxetable}{crrrrrrr}
\small
\tablecaption{Deep Multicolor Survey Fields}
\tablehead{
\colhead{Field} & 
\colhead{RA (1950.0)} & 
\colhead{Dec (1950.0)} & 
\colhead{l(II)} & 
\colhead{b(II)} & 
\colhead{Area\tablenotemark{a}} & 
\colhead{E(B-V)\tablenotemark{b}}}
\startdata
01e & 01:00:16 & $-$00:57:52 &  129.3069 &  $-$63.4266  &  292.4 & 0.0192 \nl
01w & 00:59:21 & $-$00:57:55 &  128.7997 &  $-$63.4488  &  224.9 & 0.0273 \nl
10e & 10:34:19 & $-$00:44:40 &  248.4348 &   47.1235  &  286.0 & 0.0302 \nl
14n & 13:58:40 & $-$00:37:06 &  336.7107 &   57.3613  &  288.5 & 0.0283 \nl
14s & 13:58:40 & $-$00:54:24 &  336.4456 &   57.1110  &  272.8 & 0.0263 \nl
17n & 17:14:43 & $+$50:16:47 &   76.9826 &   35.5101  &  266.0 & 0.0082 \nl
17s & 17:14:45 & $+$50:00:01 &   76.6398 &   35.4963  &  285.9 & 0.0093 \nl
21e & 21:39:57 & $-$04:00:06 &   51.8119 &  $-$39.2835  &  247.0 & 0.0243 \nl
21w & 21:39:01 & $-$04:00:01 &   51.6505 &  $-$39.0849  &  290.0 & 0.0203 \nl
22e & 22:47:04 & $-$02:09:11 &   68.6033 &  $-$51.4310  &  286.9 & 0.0542 \nl
22w & 22:46:07 & $-$02:09:14 &   68.3400 &  $-$51.2595  &  249.7 & 0.0492 \nl
\enddata
\tablenotetext{a}{Units of square arcminutes.}
\tablenotetext{b}{Calculated with a program supplied by D. Burstein, using data given in Burstein \& Heiles 1978 and Burstein \& Heiles 1982.}
\end{deluxetable}


\begin{deluxetable}{lcccccccc}
\small
\tablenum{4}
\tablecaption{Photometric Calibration Parameters}
\tablehead{
\colhead{} &
\colhead{Color} &
\colhead{Extinction} &
\colhead{1991 Base} &
\colhead{1991 Average} &
\colhead{1990 Base} &
\colhead{1990 Average} \\[.2ex]
\colhead{Filter} &
\colhead{Term} &
\colhead{Coefficient} &
\colhead{Zeropoint} &
\colhead{Zeropoint} &
\colhead{Zeropoint} &
\colhead{Zeropoint}}
\startdata
U & -0.043$\pm$0.011 &  0.530$\pm$0.023 & -20.64 & -20.69 & -20.78 & -20.71 \nl 
B & -0.063$\pm$0.016 &  0.275$\pm$0.015 & -21.71 & -21.75 & -21.80 & -21.74 \nl 
V & -0.092$\pm$0.018 &  0.113$\pm$0.007 & -22.45 & -22.50 & -22.51 & -22.46 \nl 
R & -0.098$\pm$0.020 &  0.087$\pm$0.017 & -22.06 & -22.06 & -22.16 & -22.03 \nl 
I75 & ~~~~~...~~~~~~~ &  0.078$\pm$0.014 & -22.03 & -22.05 & -22.08 & -21.99 \nl 
I86 & ~~~~~...~~~~~~~ &  0.039$\pm$0.008 & -21.48 & -21.48 & -21.53 & -21.45 \nl 
\enddata
\end{deluxetable}

\begin{deluxetable}{llccccccc}
\footnotesize
\tablecaption{Observations}
\tablehead{
\colhead{Field} & 
\colhead{Filter} & 
\colhead{Frame} & 
\colhead{Exposure\tablenotemark{a}} & 
\colhead{UT Date} & 
\colhead{UT Time} & 
\colhead{Airmass} & 
\colhead{Seeing(\arcsec)} & 
\colhead{Zeropoint\tablenotemark{b}}}
\startdata
01e & U & 1. & 1350 & 90/8/24 & 11:13  & 1.219 & 1.60 & -20.69 \nl
 & B & 1. & 1080 & 90/8/23 & 11:00  & 1.194 & 1.21 & -21.66 \nl
 & V & 1. &  540 & 90/8/23 & 10:45  & 1.177 & 1.10 & -22.37 \nl
 & R & 1. &  540 & 90/8/23 & 11:22  & 1.216 & 1.22 & -22.02 \nl
 & I75 & 1. &  540 & 90/8/23 & 11:35  & 1.239 & 1.28 & -21.98 \nl
 & I86 & 1. & 1080 & 90/8/24 & 11:41  & 1.268 & 2.03 & -21.48 \nl
01w & U & 1. & 1350 & 90/8/22 & 10:15  & 1.170 & 1.15 & -20.85 \nl
 & U & 2. & 1350 & 90/8/22 & 10:41  & 1.178 & 1.24 & -20.82 \nl
 & B & 1. & 1080 & 90/8/23 & 10:09  & 1.170 & 1.53 & -21.77 \nl
 & B & 2. & 1080 & 90/8/21 & 10:42  & 1.175 & 1.39 & -22.79 \nl
 & V & 1. &  540 & 90/8/23 & 10:31  & 1.172 & 1.26 & -22.46 \nl
 & V & 2. &  540 & 90/8/21 & 10:29  & 1.170 & 1.14 & -22.47 \nl
 & R & 1. &  540 & 90/8/22 & 9:48  & 1.181 & 0.93 & -22.02 \nl
 & R & 2. &  540 & 90/8/21 & 11:04  & 1.187 & 1.06 & -21.03 \nl
 & I75 & 1. &  540 & 90/8/21 & 10:02  & 1.173 & 1.00 & -21.96 \nl
 & I75 & 2. &  540 & 90/8/21 & 11:18  & 1.201 & 0.99 & -21.99 \nl
 & I86 & 1. & 1080 & 90/8/22 & 11:31  & 1.228 & 1.39 & -21.44 \nl
 & I86 & 2. & 1080 & 90/8/22 & 11:14  & 1.207 & 0.95 & -21.41 \nl
10e & U & 1. & 1350 & 91/4/15 & 6:07  & 1.338 & 1.36 & -20.74 \nl
 & U & 2. & 1350 & 91/4/15 & 6:43  & 1.475 & 1.37 & -20.75 \nl
 & B & 1. & 1080 & 91/4/15 & 7:10  & 1.615 & 1.29 & -21.73 \nl
 & V & 1. &  540 & 91/4/14 & 4:34  & 1.191 & 1.41 & -22.49 \nl
 & V & 2. &  540 & 91/4/14 & 4:47  & 1.195 & 1.62 & -22.48 \nl
 & R & 1. &  675 & 91/4/14 & 5:02  & 1.204 & 1.57 & -22.06 \nl
 & R & 2. &  675 & 91/4/14 & 5:17  & 1.218 & 1.66 & -22.06 \nl
 & I75 & 1. &  540 & 91/4/14 & 5:32  & 1.237 & 1.69 & -22.05 \nl
 & I75 & 2. &  540 & 91/4/14 & 5:45  & 1.258 & 1.72 & -22.05 \nl
 & I86 & 1. & 1080 & 91/4/14 & 6:00  & 1.302 & 1.65 & -21.48 \nl
 & I86 & 2. & 1080 & 91/4/15 & 5:43  & 1.244 & 1.02 & -21.50 \nl
14n & U & 1. & 1350 & 91/4/14 & 9:08  & 1.269 & 2.35 & -20.71 \nl
 & U & 2. & 1350 & 91/4/14 & 9:35  & 1.336 & 2.66 & -20.77 \nl
 & B & 1. & 1080 & 91/4/15 & 7:45  & 1.190 & 1.10 & -21.77 \nl
 & B & 2. & 1080 & 91/4/15 & 8:07  & 1.195 & 1.16 & -21.76 \nl
 & V & 1. &  540 & 91/4/14 & 7:12  & 1.210 & 1.54 & -22.50 \nl
 & V & 2. &  540 & 91/4/14 & 6:59  & 1.225 & 1.48 & -22.50 \nl
 & R & 1. &  675 & 91/4/14 & 7:26  & 1.198 & 1.74 & -22.06 \nl
 & R & 2. &  675 & 91/4/14 & 7:41  & 1.191 & 1.70 & -22.06 \nl
 & I75 & 1. &  540 & 91/4/14 & 8:09  & 1.192 & 1.95 & -22.06 \nl
 & I75 & 2. &  540 & 91/4/14 & 7:56  & 1.189 & 2.03 & -21.06 \nl
\tablebreak
 & I86 & 1. & 1080 & 91/4/14 & 8:23  & 1.202 & 1.87 & -21.48 \nl
 & I86 & 2. & 1080 & 91/4/14 & 8:45  & 1.225 & 1.91 & -21.48 \nl
14s & U & 1. & 1350 & 91/4/16 & 9:59  & 1.459 & 1.69 & -20.61 \nl
 & U & 2. & 1350 & 91/4/16 & 8:45  & 1.244 & 1.43 & -20.57 \nl
 & B & 1. & 1080 & 91/4/15 & 7:10  & 1.208 & 1.07 & -21.77 \nl
 & B & 2. & 1080 & 91/4/17 & 9:48  & 1.420 & 1.88 & -21.77 \nl
 & V & 1. &  540 & 91/4/16 & 8:01  & 1.196 & 1.31 & -22.30 \nl
 & V & 2. &  540 & 91/4/15 & 8:51  & 1.238 & 1.02 & -22.50 \nl
 & R & 1. &  675 & 91/4/16 & 8:31  & 1.203 & 1.40 & -21.33 \nl
 & R & 2. &  675 & 91/4/15 & 9:05  & 1.263 & 1.08 & -22.06 \nl
 & I75 & 1. &  540 & 91/4/17 & 8:31  & 1.221 & 2.12 & -22.03 \nl
 & I75 & 2. &  540 & 91/4/16 & 8:31  & 1.216 & 1.44 & -21.71 \nl
 & I86 & 1. & 1080 & 91/4/14 & 6:00  & 1.297 & 1.17 & -21.46 \nl
 & I86 & 2. & 1080 & 91/4/16 & 9:35  & 1.259 & 1.27 & -21.44 \nl
17n & U & 1. & 1350 & 91/4/16 & 10:47  & 1.053 & 1.57 & -20.38 \nl
 & U & 2. & 1350 & 91/4/15 & 11:01  & 1.053 & 1.06 & -20.65 \nl
 & B & 1. & 1080 & 91/4/15 & 10:37  & 1.056 & 0.97 & -21.73 \nl
 & B & 2. & 1080 & 91/4/16 & 11:14  & 1.055 & 1.40 & -21.52 \nl
 & V & 1. &  540 & 91/4/14 & 10:21  & 1.066 & 1.47 & -22.50 \nl
 & V & 2. &  540 & 91/4/14 & 10:36  & 1.059 & 1.47 & -22.50 \nl
 & R & 1. &  675 & 91/4/14 & 10:50  & 1.055 & 1.34 & -22.06 \nl
 & R & 2. &  675 & 91/4/14 & 11:06  & 1.053 & 1.25 & -22.06 \nl
 & I75 & 1. &  540 & 91/4/14 & 11:21  & 1.054 & 1.20 & -22.06 \nl
 & I75 & 2. &  540 & 91/4/14 & 11:33  & 1.056 & 1.16 & -22.06 \nl
 & I86 & 1. & 1080 & 91/4/15 & 11:29  & 1.058 & 0.90 & -21.48 \nl
 & I86 & 2. & 1080 & 91/4/15 & 10:14  & 1.065 & 0.90 & -21.48 \nl
17s & U & 1. & 1350 & 90/8/21 & 5:42  & 1.289 & 1.27 & -20.76 \nl
 & U & 2. & 1350 & 90/8/24 & 4:20  & 1.137 & 1.26 & -20.75 \nl
 & B & 1. & 1080 & 90/8/21 & 4:53  & 1.169 & 1.39 & -21.79 \nl
 & B & 2. & 1080 & 90/8/23 & 4:04  & 1.107 & 1.44 & -21.77 \nl
 & V & 1. &  540 & 90/8/21 & 4:40  & 1.139 & 1.11 & -22.49 \nl
 & V & 2. &  540 & 90/8/23 & 4:26  & 1.129 & 1.15 & -22.49 \nl
 & R & 1. &  540 & 90/8/23 & 4:45  & 1.159 & 1.21 & -22.02 \nl
 & R & 2. &  540 & 90/8/21 & 5:16  & 1.205 & 1.23 & -22.05 \nl
 & I75 & 1. &  540 & 90/8/21 & 5:29  & 1.234 & 1.06 & -21.99 \nl
 & I75 & 2. &  540 & 90/8/23 & 4:57  & 1.183 & 1.35 & -21.95 \nl
 & I86 & 1. & 1080 & 90/8/23 & 3:41  & 1.083 & 1.25 & -21.40 \nl
 & I86 & 2. & 1080 & 90/8/23 & 6:10  & 1.371 & 1.30 & -21.41 \nl
\tablebreak
21e & U & 1. & 1350 & 90/8/24 & 5:27  & 1.318 & 1.71 & -20.83 \nl
 & U & 2. & 1350 & 90/8/24 & 5:53  & 1.271 & 1.48 & -20.83 \nl
 & B & 1. & 1080 & 90/8/23 & 5:57  & 1.274 & 1.25 & -21.81 \nl
 & B & 2. & 1080 & 90/8/24 & 6:44  & 1.233 & 1.78 & -21.81 \nl
 & V & 1. &  540 & 90/8/24 & 7:06  & 1.234 & 1.54 & -22.50 \nl
 & V & 2. &  540 & 90/8/23 & 5:42  & 1.308 & 1.28 & -22.49 \nl
 & R & 1. &  540 & 90/8/24 & 7:19  & 1.240 & 1.57 & -22.03 \nl
 & R & 2. &  540 & 90/8/23 & 6:20  & 1.251 & 1.21 & -22.02 \nl
 & I75 & 1. &  540 & 90/8/24 & 7:32  & 1.250 & 1.38 & -22.00 \nl
 & I75 & 2. &  540 & 90/8/23 & 6:33  & 1.241 & 1.34 & -21.99 \nl
 & I86 & 1. & 1080 & 90/8/23 & 7:13  & 1.237 & 1.11 & -21.44 \nl
 & I86 & 2. & 1080 & 90/8/24 & 6:21  & 1.243 & 1.28 & -21.44 \nl
21w & U & 1. & 1350 & 90/8/21 & 7:41  & 1.256 & 1.86 & -20.66 \nl
 & U & 2. & 1350 & 90/8/22 & 6:23  & 1.246 & 1.44 & -20.72 \nl
 & B & 1. & 1080 & 90/8/21 & 6:51  & 1.234 & 1.86 & -21.77 \nl
 & B & 2. & 1080 & 90/8/22 & 5:34  & 1.321 & 1.19 & -21.77 \nl
 & V & 1. &  540 & 90/8/21 & 6:38  & 1.242 & 1.55 & -22.49 \nl
 & V & 2. &  540 & 90/8/22 & 5:21  & 1.366 & 1.17 & -22.48 \nl
 & R & 1. &  540 & 90/8/22 & 5:57  & 1.285 & 1.02 & -22.02 \nl
 & R & 2. &  540 & 90/8/21 & 7:15  & 1.234 & 1.56 & -22.06 \nl
 & I75 & 1. &  540 & 90/8/21 & 7:28  & 1.239 & 1.52 & -22.01 \nl
 & I75 & 2. &  540 & 90/8/22 & 6:09  & 1.266 & 1.00 & -21.99 \nl
 & I86 & 1. & 1080 & 90/8/21 & 6:08  & 1.291 & 1.64 & -21.48 \nl
 & I86 & 2. & 1080 & 90/8/22 & 6:50  & 1.233 & 1.54 & -21.46 \nl
22e & U & 1. & 1350 & 90/8/24 & 8:27  & 1.218 & 1.80 & -20.62 \nl
 & U & 2. & 1350 & 90/8/24 & 8:54  & 1.248 & 1.38 & -20.66 \nl
 & B & 1. & 1080 & 90/8/24 & 9:42  & 1.349 & 2.32 & -21.68 \nl
 & B & 2. & 1080 & 90/8/23 & 8:30  & 1.215 & 1.06 & -21.69 \nl
 & V & 1. &  540 & 90/8/23 & 8:15  & 1.206 & 1.09 & -22.44 \nl
 & V & 2. &  540 & 90/8/24 & 10:04  & 1.407 & 1.86 & -22.45 \nl
 & R & 1. &  540 & 90/8/23 & 8:51  & 1.230 & 0.98 & -22.07 \nl
 & R & 2. &  540 & 90/8/24 & 10:17  & 1.461 & 2.05 & -22.00 \nl
 & I75 & 1. &  540 & 90/8/23 & 9:04  & 1.246 & 1.11 & -22.01 \nl
 & I75 & 2. &  540 & 90/8/24 & 10:30  & 1.522 & 1.83 & -22.03 \nl
 & I86 & 1. & 1080 & 90/8/24 & 9:21  & 1.293 & 1.63 & -21.48 \nl
 & I86 & 2. & 1080 & 90/8/23 & 9:44  & 1.343 & 1.09 & -21.46 \nl
\tablebreak
22w & U & 1. & 1350 & 90/8/24 & 7:59  & 1.206 & 1.57 & -20.52 \nl
 & U & 2. & 1350 & 90/8/22 & 9:06  & 1.257 & 1.23 & -20.46 \nl
 & B & 1. & 1080 & 90/8/22 & 8:18  & 1.207 & 1.24 & -21.62 \nl
 & B & 2. & 1080 & 90/8/21 & 9:09  & 1.250 & 1.59 & -21.67 \nl
 & V & 1. &  540 & 90/8/21 & 8:55  & 1.227 & 1.52 & -22.45 \nl
 & V & 2. &  540 & 90/8/22 & 8:05  & 1.205 & 1.12 & -22.42 \nl
 & R & 1. &  540 & 90/8/22 & 8:40  & 1.217 & 1.01 & -22.02 \nl
 & R & 2. &  540 & 90/8/21 & 9:40  & 1.302 & 1.45 & -22.04 \nl
 & I75 & 1. &  540 & 90/8/22 & 8:53  & 1.229 & 1.07 & -21.98 \nl
 & I75 & 2. &  540 & 90/8/21 & 9:53  & 1.334 & 1.54 & -22.00 \nl
 & I86 & 1. & 1080 & 90/8/23 & 7:47  & 1.207 & 1.11 & -21.47 \nl
 & I86 & 2. & 1080 & 90/8/22 & 7:14  & 1.234 & 1.57 & -21.47 \nl
\enddata
\tablenotetext{a}{Units of seconds.}
\tablenotetext{b}{See \S 5.3.}
\end{deluxetable}


\begin{deluxetable}{cccccc}
\small
\tablecaption{Average Magnitude Limits}
\tablehead{
\colhead{} &
\colhead{Threshold} &
\colhead{5$\sigma$~Limiting} &
\colhead{3$\sigma$~Limiting} &
\colhead{Upper Limit} &
\colhead{} \\[.2ex]
\colhead{Filter} &
\colhead{Magnitude} &
\colhead{Magnitude\tablenotemark{a}} &
\colhead{Magnitude\tablenotemark{b}} &
\colhead{Magnitude} &
\colhead{G(m$_{5\sigma}$)\tablenotemark{c}}}
\startdata
U & 21.0 & 22.8 & 23.45 & 16.0 & 25.2\% \nl
B & 21.5 & 23.8 & 24.30 & 16.5 & 22.2\% \nl
V & 21.0 & 23.5 & 24.00 & 16.5 & 27.8\% \nl
R & 20.0 & 23.0 & 23.50 & 16.5 & 33.0\% \nl
I75 & 20.0 & 22.4 & 23.05 & 16.0 & 19.5\% \nl
I86 & 19.5 & 22.1 & 22.70 & 16.0 & 21.5\% \nl
\enddata
\tablenotetext{a}{Also 90\% completeness magnitude.}
\tablenotetext{b}{Also 50\% completeness magnitude.}
\tablenotetext{c}{The percentage of misclassified galaxies in the stellar
object catalog at the 5$\sigma$\ limiting magnitude.}
\end{deluxetable}

\begin{figure}
\plotone{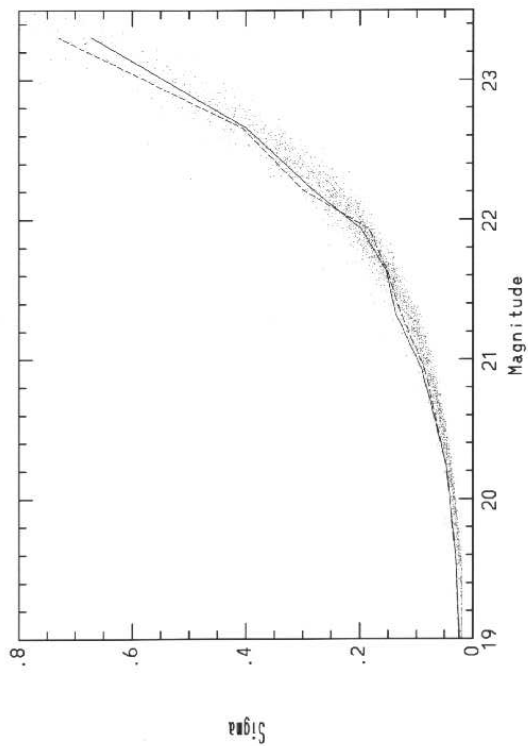}
\caption{
Formal photometric errors vs. errors estimated from the magnitude
difference plots like that shown in Figure 3.  The points are the photometric
errors from the IRAF `phot' routine for all objects in both CCD frames in the
21w field, I8600 filter.  The dashed line is the sigma estimated from the
average square of the magnitude difference for every 50 objects, while the
connected line is the sigma estimated for every 50 objects using the average
absolute value of the magnitude difference.
The match is clearly quite good at all magnitudes.
}
\end{figure}

\end{document}